\documentclass{amsart}
\usepackage{graphicx}
\usepackage[T2A]{fontenc}
\usepackage[cp1251]{inputenc}
\usepackage[english, russian]{babel}
\usepackage{amsfonts,amssymb,mathrsfs,amsmath,amsbsy,amsthm,amscd}

\newtheorem{proposition}{Proposition}
\begin{document}


\author{M.\,A.~Martynov, O.\,S.~Rozanova}

\title[A certain estimate of
volatility through return]{A certain estimate of volatility through
return for stochastic volatility models}

\maketitle



\begin{abstract}
We study the dependence of volatility on the stock price in the
stochastic volatility  framework on the example of  the Heston
model.
 To be more specific, we consider the conditional
 expectation of variance (square of volatility) under fixed stock price return as
a function of the  return  and time. The  behavior of this function
depends on the initial  stock price return distribution density. In
particular,  we show that the graph of the conditional
 expectation of variance is convex downwards
near the mean value of the stock price return. For the Gaussian
distribution this effect is strong, but it weakens and becomes
negligible as the decay of distribution at infinity slows down.
\bigskip

\end{abstract}

\section{ Introduction}

Stochastic volatility (SV) models are quite popular in recent
decades due to a need for reliable quantitative analysis of market
data. The most popular ones are the Heston \cite{Heston},
Stein-Stein \cite{Stein}, Sch$\rm \ddot{o}$ble-Zhu \cite{Schobel},
Hull-White \cite{Hull} and Scott \cite{Scott} models. We refer for
reviews to \cite{Micciche}, \cite{Mitra}, \cite{Fouque}. The main
reason for introducing the SV models is to find a realistic
alternative approach to option pricing to capture the time varying
nature of the volatility, assumed to be constant in the
Black-Scholes approach.

Nevertheless, SV models can be used for investigation of another
properties of financial markets. For example, in \cite{Dragulescu}
the time-dependent probability distribution of stock price returns
was studied. While returns are readily known from a financial data,
variance (square of the stock-price volatility) is not given
directly, so it acts as a hidden stochastic variable. In
\cite{Dragulescu}  the joint probability density function of returns
and variance  was found, then the integration over variance was
performed  and the probability distribution function of returns
unconditional on variance was obtained. The latter PDF can be
directly compared with the Dow-Jones data for the 20-years period of
1982 - 2001 and  an excellent agreement was found. The tails of the
PDF decay slower than the log-normal distribution predicts (the
so-called "fat-tails" \,effect).

 Technically our paper is connected with \cite{Dragulescu}. However, we study
 the dependence of the variance on
fixed returns, thus, we estimate hidden stochastic variable through
the variable that can be easily obtained from  financial data. The
result strongly depends on initial distribution of returns and
variance. It is natural that the distributions change their shape
with time. In particular, we show that for Gaussian initial
distribution of returns the expectation of variance demonstrates the
convexity downwards  near the mean value of returns.

\section{General formulas for the conditional expectation and variance}


Let us consider the stochastic differential equation system:
\begin{equation}
\label{systgen}
\begin{array}{ll}
dF_t = A dt + \sigma dW_{1}, \quad dV_t = B dt + \lambda dW_{2}, \\
F_0 = f, \quad V_0 = v, \quad t\ge 0, \, f \in \mathbb R, v
\in\mathbb R,
\end{array}
\end{equation}
where $W(t)=(W_{1}(t),W_{2}(t)) $ is a two-dimensional standard
Wiener process, $A=A(t, F_t, V_t),\; B=B(t, F_t, V_t),\;
\sigma=\sigma(t, F_t, V_t),\; \lambda=\lambda(t, F_t, V_t)$ are
prescribed functions.

The joint probability density $P(t,f,v)$ of random values $F_t$ and
$V_t$ obeys the Fokker--Plank equation (e.g.,\cite{Risken})
\begin{equation}
\label{f-peq}
 \frac{\partial P}{\partial t} = - \frac{\partial}{\partial f}\left( AP\right)
 - \frac{\partial}{\partial v}\left( BP\right)
 + \frac{1}{2}\frac{\partial^2}{\partial f^2}\left( \sigma^2 P\right)
 +\frac{1}{2}\frac{\partial^2}{\partial v^2}\left( \lambda^2 P\right)
\end{equation}
with initial condition
\begin{equation}
\label{f-pzero}
P(0,f,v)=P_{0}(f,v),
\end{equation}
determined by initial distributions of $F_t$ and $V_t$.

If $P(t,f,v)$ is known, one can find $E \left(V_t|F_t=f\right)$,
which is the conditional  expectation of value $V_t$ at a fixed
$F_t$ at the moment $t$. This value can be found by the following
formula (see, \cite{Chorin}):
\begin{equation}
\label{condexp} E \left(V_t|F_t=f\right) =
\lim_{L\to+\infty}\,\frac{\int_{(-L,L)} v P(t,f,v)dv}{\int_{(-L,L)}
P(t,f,v)dv}.
\end{equation}
 Let
us also define the variance of  $V_t$ at a fixed $F_t$ as
\begin{equation}
\label{disp} Var \left(V_t|F_t=f\right)
=\lim_{L\to+\infty}\,\frac{\int_{(-L,L)} v^2
P(t,f,v)dv}{\int_{(-L,L)} P(t,f,v)dv}- E^2 \left(V_t|F_t=f\right).
\end{equation}
In this both formulae the improper integrals  from numerator and
denominator are assumed to converge. The assumption imposes a
restriction on the coefficients $A, B, \sigma, \lambda$.

Note that if we choose $P_{0}(f,v)=\delta (v-v_{0}(f))g(f)$, where
$v_{0}(f)$ and $g(f)$ are arbitrary smooth functions, then $E
\left(V_t|F_t=f\right)|_{t=0} = v_{0}(f)$.

For some classes of  systems \eqref{systgen} the conditional
expectation $V(t,f)$ was found in \cite{Risken},\cite{AR_MMMAS},
\cite{AR_PAMS} within an absolutely different context.

Let us remark that sometimes it is easier to find the Fourier
transform of $P(t,f,v)$ function over $f, v$ variables, than the
function itself. We will get formula allowing to express $E
\left(V_t|F_t=f\right)$  in terms of Fourier transform of $P(t,f,v)$
and will apply it  for finding an average variance  of the stock
price, which depends on known return rate.
%


\begin{proposition}\label{lem1}
Let  $\hat P(t, \mu, \xi)$ be the Fourier transform of function
$P(t, f, v)$ over $(f, v)$ variables, which is the solution of
problem \eqref{f-peq},~\eqref{f-pzero}, and both integrals from
\eqref{condexp} converge. Assume that $\hat P (t, \mu, 0)$ and
$\partial_\xi \hat P(t, \mu, 0)$ are decreasing over $\mu$ at
infinity faster than any power. Then $E \left(V_t|F_t=f\right)$ and
$Var \left(V_t|F_t=f\right)$ determined by \eqref{condexp} and
\eqref{disp} can be found as
\begin{equation}
\label{condexpth} E \left(V_t|F_t=f\right) =\frac{i {\bf F}^{-1}_\mu
\, [\partial_\xi \hat P(t,\mu,0)](t,f) } {{\bf F}^{-1}_\mu \,  [
\hat P(t,\mu,0)](t,f) }, \quad t\ge 0, \, f\in \mathbb{R},
\end{equation}
\begin{equation}
\label{pd1} Var \left(V_t|F_t=f\right) =\frac{({\bf F}^{-1}_\mu
[\partial_\xi \hat P(t,\mu,0)])^2-{\bf F}^{-1}_\mu [\partial^2_\xi
\hat P(t,\mu,0)] {\bf F}^{-1}_\mu [\hat P(t,\mu,0)]} {( {\bf
F}^{-1}_\mu [\hat P(t,\mu,0)])^2}(t,f),
\end{equation}
where ${\bf F}^{-1}_\mu$ and ${\bf F}^{-1}_\xi$ mean the inverse
Fourier transforms over $\mu$ and $\xi$, respectively.
\end{proposition}

The proof is a simple exercise in the Fourier analysis.

\section{Example: the Heston model}

Of course, there is no explicit formula for the joint probability
density function $P(t,f,v)$ for arbitrary system \eqref{systgen}. We
will consider a particular, but important case of the Heston model
\cite{Heston}:
\begin{equation}
\label{heston1} df_t = \left(\alpha - \frac{v_t}{2}\right) dt +
\sqrt{v_t} dW_1 ,
\end{equation}
\begin{equation}
\label{heston2}
dv_t = -\gamma(v_t - \theta)dt + k \sqrt{v_t} dW_2 .
\end{equation}

Here $\alpha, \,\gamma, \; k, \; \theta$ are arbitrary positive
constants.

Equation \eqref{heston2} describes the process that in financial
literature is called Cox-Ingersoll-Ross (CIR) process, and in
mathematical statistics --- the Feller process \cite{Fouque},
\cite{Feller}. In \cite{Feller} it is shown that this equation has a
nonnegative solution for $t\in [0,+\infty)$ when $ 2\gamma \theta>
k^2$.

The first equation describes a  return $f_t$ on the stock price, in
assumption  that the stock price itself obeys a geometric Brownian
motion with stochastic volatility. The second equation describes the
square  of  volatility $\sigma^{2}_t = v_t$.


The Fokker--Planck equation \eqref{f-peq} for the joint density
function $P(t, f, v)$ of return $f_t$ and variance $v_t$  takes here
the following form:

\begin{equation*}
 \frac{\partial P(t,f,v)}{\partial t} =
 \gamma P(t,f,v) + (\gamma(v - \theta) + k^2) \frac{\partial P(t,f,v)}{\partial v}
 + \left(\frac{v}{2} - \alpha\right) \frac{\partial P(t,f,v)}{\partial f}
 +
\end{equation*}
\begin{equation}
\label{f-peq2}
\frac{k^2 v}{2} \frac{\partial^2 P(t,f,v)}{\partial
v^2}
 + \frac{v}{2}\frac{\partial^2 P(t,f,v)}{\partial f^2}.
\end{equation}
Now we can choose different initial distributions for return and
variance. Note that it is natural to assume that initially the
variance does not depend on return. 

Below we denote $E\left(v_t| f_t=f\right)$  as $V(t,f)$ for short.

The function $\hat P \left( t,\mu,\xi \right)$,   the Fourier
transform of $P$ over $(f, v)$,  satisfies the equation
\begin{equation}
\label{f-peqfur}
 \frac{\partial \hat P \left( t,\mu,\xi \right)}{\partial t}
 + \frac12 \left(\mu + i\mu^2 + 2\gamma \xi + i k^2 \xi^2\right) \frac{\partial \hat P \left( t,\mu,\xi \right)}{\partial \xi}
 + i\left(\gamma \theta + \xi \mu \alpha \right) \hat P \left( t,\mu,\xi \right) =
 0.
\end{equation}
The first-order PDE \eqref{f-peqfur}  can be integrated, the
solution has the following form:
\begin{equation}\label{gen_sol}
{\hat P}( t,\mu,\xi) ={e^{{\frac { \left(
-i\mu\,\alpha\,{k}^{2}+{{\gamma}}^{2}{\theta}
 \right)\,t }{{k}^{2}}}}}
 \left( \frac {{k}^{2} \left( i\,(2\,\gamma\,\xi+\mu)-{\mu}^{2}-{k}^{
2}{\xi}^{2} \right) }{q}
 \right)^{-{\frac {\gamma\,\theta}{k^2}}}
\end{equation}
\begin{equation}
\ast\,F \left( \mu,- t + \frac{2\,i\arctan \left( \frac
{-k^{2}\xi+i\gamma}{ \sqrt q}\right)}
 {\sqrt {q}}\right) \,,
\end{equation}
where $q=-i{k}^{2}\mu+{k}^{2}{\mu}^{2}+{{\gamma}}^{2}$, $F$ is an
arbitrary differentiable function of two variables.

\subsection{The uniform initial distribution of returns}

We begin with the simplest and almost trivial case. Let us assume
that initially the rate of return is distributed uniformly in the
interval $(-L,L),$  ($L={\rm const}>0$), and volatility is equal to
some constant $a\ge 0$. Then the initial joint density distribution
of $f_t$ and $v_t$ is
\begin{equation}
\label{f-pzero2}
P(0,f,v)=\frac{1}{2L} \delta(v-a).
\end{equation}
To simplify further calculations we will exclude randomness for
$t=0$, i.e. we will assume  $a=0$.

 The respective initial condition for the Fourier transform is
\begin{equation}
\label{f-pzero3} \hat P(0,\mu,\xi)=\frac{\pi}{L} \delta(\mu).
\end{equation}

 The
solution of problem \eqref{f-peqfur}, \eqref{f-pzero3} takes the
form
\begin{equation}
\label{f-psolution} \hat P \left( t,\mu,\xi \right) = \frac{\pi}{L}
\delta(\mu) \left(\frac{4\gamma^2 e^{2\gamma t}} { \left( 2\gamma
e^{\gamma t} + i k^2 \xi \left(e^{\gamma t} - 1\right) \right)^2 }
\right) ^{\frac{\gamma \theta t}{k^2}}
\end{equation}

It is easy to calculate that
\begin{equation}
\label{f-psolution2}
\hat P(t,\mu,0)= \frac{\pi}{L} \delta(\mu), \quad
\partial_\xi \hat P(t,\mu,0) = \frac{\pi}{L} \delta(\mu) i \theta \left(e^{-\gamma t} -
1\right).
\end{equation}

Finally from \eqref{condexpth} and \eqref{pd1} we get
\begin{equation}
\label{f-psolution3} E\left(v_t| f_t=f\right) = \theta \left(1 -
e^{-\gamma t}\right),
\end{equation}
\begin{equation}
 Var \left(v_t|f_t=f\right) = \frac{\theta k^2}{2 \gamma} \left(1 -
e^{-\gamma t}\right)^2.
\end{equation}

It is evident that here there is no dependence on $f$ and the result
is the same as we could obtain from calculation of mathematical
expectation and variance of $v_t$ from equation \eqref{heston2}.

\subsection {The Gaussian initial distribution of returns}

Let us assume that initially rate of return is distributed according to the Gaussian law.
Then we have the following initial condition:
\begin{equation}
\label{gf-pzero2} P(0,f,v) = \frac{m}{\sqrt{\pi}}e^{-m^2 f^2}
\delta(v),\quad m>0.
\end{equation}
When $a=0$, the Fourier transform of initial data over $(f,v)$ is
$\hat P \left( 0,\mu,\xi \right)= e^{-\frac{\mu^2}{4 m^2}}$.

Solution of the problem \eqref{f-peqfur}, \eqref{gf-pzero2} takes
the form:
\begin{equation}\label{pr_fur}
\hat
P(t,\mu,\xi)=\frac{\sqrt{\pi}}{m}\left(-\frac{\mu(\mu-i)+\frac{\gamma^2}{k^2}}{\mu^2+k^2\gamma^2-i(2\gamma\xi+\mu)}\right)
^\frac{\gamma\theta}{k^2}\,\exp\left(-\frac{\mu^2}{4m^2}-(\alpha\mu
i-\frac{\gamma^2\theta}{k^2})t\right)\ast\, \end{equation}
\begin{equation*}
\\\left(-\cosh\left(\frac{t}{2}\sqrt{k^2\mu(\mu-i)+\gamma^2}-
i\arctan\left(\frac{-k^2\xi+i\gamma}{\sqrt{k^2\mu(\mu-i)+\gamma^2}}\right)\right)\right)^{-\frac{2\gamma\theta}{k^2}}.
\end{equation*}

We see that $\hat P(t,\mu,\xi)$ exponentially decreases over $\mu$.
That is why we can use formula \eqref{condexpth} and obtain  (after
cumbersome transformations) the following integral expression:
\begin{align}
V(t,f) = 2 \gamma \theta\,\frac{\int\limits_{\mathbb R} \Phi(t,\mu,
f) d\mu}
  {\int\limits_{\mathbb R} \Psi(t,\mu,f) d\mu}
 \label{gauss1},
\end{align}
where
$$
\Psi(t,\mu,f)= e^{\frac{ - \mu^2 +  i\mu  (4f - 4t\alpha-1) }{4
m^2}}\,
\left(\frac{\lambda}{\left({\lambda}\cosh{\left(\frac{{\lambda}t}{4}
\right)}
 +2 \gamma \sinh{\left(\frac{{\lambda}t}{4} \right)}\right)}\right)^{\frac{2\gamma
 \theta}{k^2}},\,$$
$$\Phi(t,\mu, f)=\Psi(t,\mu, f)\frac{\sinh{\left(\frac{\lambda t}{4} \right)}}
 {\left({\lambda} \cosh{\left(\frac{ {\lambda} t}{4}\right)}
 +2 \gamma \sinh{\left(\frac{{\lambda}t}{4} \right)} \right)},$$
 $$
 \lambda^2=k^2(4\mu^2+1)+4
\gamma^2.$$

Let us remark that if $a\ne 0$, we can also get a similar formula,
but it will be more cumbersome.

The limit case as $m\to\infty$ for \eqref{gf-pzero2} is
\begin{equation}
\label{gf-pzezo-delta} P(0,f,v) = \delta(f) \delta(v),\quad m>0.
\end{equation}
For this case the formula \eqref{gauss1} modifies as follows: the
exponential factor in the expression for $\Psi$ takes the form $e^{
i\mu  (f - t\alpha) }$.

\subsection{``Fat-tails'' initial distribution of returns}

Integral formula, analogous to \eqref{gauss1} can be obtained for
initial distributions intermediate between uniform and Gaussian
ones. For example, as initial distribution we can take
$$P(0,f,v) =
K(1+m^2 f^2)^q \, \delta(v),\quad m>0,\,q<0,$$ with an appropriate
constant $K$. Exact formula for the Fourier transform $\hat
P(t,\mu,\xi)$ can be found for $q=-\frac{1}{2}, \,-n,\,n\in\mathbb
N.$ For all these cases $\hat P(t,\mu,\xi)$ decays as $|\mu| \to
\infty$ sufficiently fast and Proposition \ref{lem1} can be applied
for calculation of $V(t,f)$.

For example, for $q=-1$ the difference with \eqref{pr_fur} is only
in the multiplier $e^{-\frac{\mu^2}{4m^2}}$: it should be changed to
$$\left({e^{{-\frac {\mu}{m}}}} -{e^{{\frac {\mu}{m}}}} \right) {\it
H} \left( \mu \right) +{e^{{\frac {\mu}{m}}}},$$ with the Heaviside
function $H$.

\subsection{Convexity downward of the volatility curve and asymptotic behavior for
small time}

It turns out that if in the Heston model the average volatility is
considered as a function of the rate on return, we will observe a
deflection of the plot.
The effect appears in numerical calculation of both integrals in
\eqref{gauss1} with the use of standard algorithms. The numerical
calculation  of the integrals
 over an infinite  interval is based on the QUADPACK routine QAGI \cite{Piessens},
 where the entire infinite integration range is first transformed to
 the segment $ [0,1]$. For example, Fig.~1 presents  the graph of function
$V(t,f)$ at three consequent moments of time  for the following
values of parameters:
$
\gamma=1,\,k=1,\,\theta=1,\,\alpha=1,\,m=1.
$

This behavior of the volatility plot can be studied  by analytical
methods as well. Indeed, let us fix rate of return $f$. Then from
\eqref{gauss1} by expansion of integrand functions into formal
series as $t \rightarrow 0$ up to the forth component and by further
term-wise integration (series converge at least for small $f$ and
$m$) we will get that
\begin{equation}\label{expand}
\begin{array}{l}
V(t,f) = \gamma\theta t - \frac12 \gamma^2 \theta t^2 + \frac16
\gamma \theta \left(\gamma^2 + 2f^2 m^4 k^2 - f m^2 k^2 - m^2
k^2\right)t^3 \\ - \frac16 \gamma \theta \left(8
\gamma\,{k}^{2}{f}^{2}{m}^{4}-4 \left( {\gamma}\,+4 \,{m}^{2}\alpha
\right) {k}^{2}{m}^{2}f-4 \left( \gamma+\alpha
 \right) {m}^{2}{k}^{2}+{\gamma}^{3}
      \right)\,t^4+ O(t^5).
\end{array}
\end{equation}
Let us justify a possibility to expand $V(t,f)$ into the Taylor
series.  We should prove that both integrals in the numerator and
denominator of \eqref{gauss1} can be differentiated with respect to
$t$. Indeed, let $t\in \Omega_t=(-\tau,\tau),\,0<\tau<\infty$. It
can be readily shown that both integrands in \eqref{gauss1}, $\Phi$
and $\Psi$, are continuous with respect to $\mu$ and $t$ on $\mathbb
R\times \Omega_t$, the derivatives of any order
$\partial_t^n\,\Phi$, $\partial_t^n\,\Psi$, $n=0,1,...$ are also
continuous on $\mathbb R\times \Omega_t$. Moreover,
$|\partial_t^n\,\Phi|$, $|\partial_t^n\,\Psi|$ can be estimated from
above by $c_1\cdot e^{-c_2 \mu^2}$, with positive constants $c_1$
and $c_2$. Therefore $\int_{\mathbb R}\,\partial_t^n\,\Phi\, d\mu $
and $\int_{\mathbb R}\,\partial_t^n\,\Psi\, d\mu $ converge
uniformly on $\Omega_t$. Thus, according to the classical theorem of
calculus the  numerator and denominator in  \eqref{gauss1} can be
differentiated on $\Omega_t$ under the integral sign. Since for
$$\partial_t^n\,\int_{\mathbb
R}\,Q(t,\mu,f)\, d\mu|_{t=0}\,=\,\int_{\mathbb
R}\,\partial_t^n\,Q(t,\mu,f)|_{t=0}\, d\mu,$$ $Q=\Psi$ or $\Phi$,
 the Taylor
coefficients in the expansion of $\int_{\mathbb R}\,Q(t,\mu,f)\,
d\mu$  can be obtained by integration  of the respective coefficient
of $Q(t,\mu,f)$ of the Taylor series in $t$ with respect to $\mu$.
The latter integrals can be explicitly calculated. This gives
expansion \eqref{expand}.

Hence for  $t\to 0$ we find that
\begin{equation*}
\begin{array}{l}
V(t,f) \sim \frac13\,{\gamma}\,{\theta}\,{t}^{3}(1-{\gamma}\,{t})
{m}^{4}{k}^{2}{f}^{2}- \left(
\frac23\,\,\alpha\,{m}^{2}\,{t}+\frac16 \left(1\,- {{\gamma}}\,{t}
\right)
 \right){t}^{3}{\gamma}\,{\theta}{m}^{ 2} {k}^{2}f+ \\\frac16\,\left({{\gamma}}\,{t}
- \left(1-\alpha\,{t} \right)  \right){\theta}\,{\gamma}{t}^{3}
{m}^{2}{k}^{2}+{\gamma}\,{\theta}\,t-\frac12\,{{\gamma}}
^{2}{\theta}\,{t}^{2}+\frac16\,{{\gamma}}^{3}{\theta}\,{t}^{3}-\frac1{24}
\,{{\gamma}}^{4}{\theta}\,{t}^{4}
\end{array}
\end{equation*}
is a quadratic trinomial over $f$ with a minimum in point $f={\frac
{4\,{m}^{2}\alpha\,t-\gamma\,t+1}{4\,{m}^{2} \left( 1-\gamma\,t
 \right) }},
$ for $t>0$.

The  effect holds for initial ``fat-tails'' power initial
distributions as well. Nevertheless, this effect weakens as the
decay of the distribution at infinity becomes slower.

Fig.2 presents the function $V(t,f)$ for three consequent moments of
time for the initial distribution of return $p(f)$ given by formula
$$p(f) =\frac{1}{\pi}\frac
{1}{1+f^2}.$$ The  values of parameters are
$
\gamma=10,\,k=1,\,\theta=0.1,\,\alpha=10.
$
 It seems
that the curves are strait lines, but the analysis of numerical
values shows that the deflection still persists near the mean value
of return. Acting as in the case of the Gaussian initial
distribution one can find the Taylor expansion of $V(t,f)$ as $t\to
0$,
$$
V(t,f) = \gamma\theta t - \frac12 \gamma^2 \theta t^2 -\gamma \theta
\frac{R_4(f,\gamma,k)}{R_6(f)} t^3+\gamma^2\theta
\frac{R_8(f,\gamma,k,\alpha)}{R_4(f)} t^4+ O(t^5),
$$
where we denote by $R_k$ a polynomial of order $k$ with respect to
$f$. We do not write down these polynomial, let us only note that
$\frac{R_4(f,\gamma,k)}{R_6(f)}\sim \frac{1}{2f^2
}\left(\frac{k^2}{8}-\gamma^2\right)$ and
$\frac{R_8(f,\gamma,k,\alpha)}{R_4(f)}\sim \frac{16f^4}{3}
\left(2\gamma^2-k^2\right)$ as ${|f|\to\infty}$.

It is very interesting to study the asymptotic behaviour of $V(t,f)$
as $|f|\to \infty$ and $t\to\infty$. We do not dwell here on this
quite delicate question at all and  reserve it for future research.
Some hints can be found in \cite{Dragulescu}, \cite{Gulisashvili1},
\cite{Gulisashvili2}.
\begin{figure}
\begin{minipage}{0.5\columnwidth}
\centerline{\includegraphics[width=0.7\columnwidth]{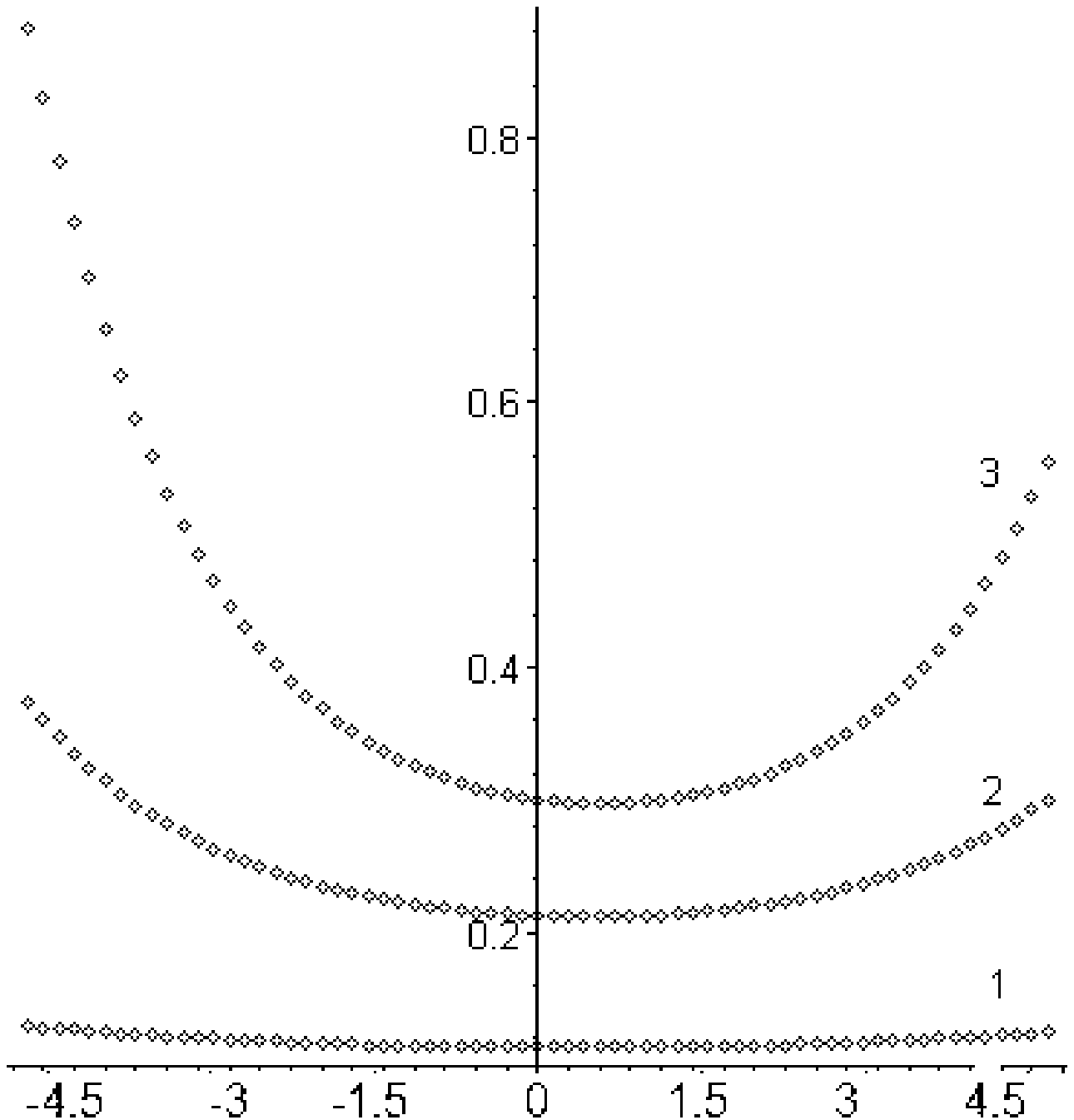}\label{smile1}}
  \caption{}
\end{minipage}%
\begin{minipage}{0.5\columnwidth}
\centerline{\includegraphics[width=0.7\columnwidth]{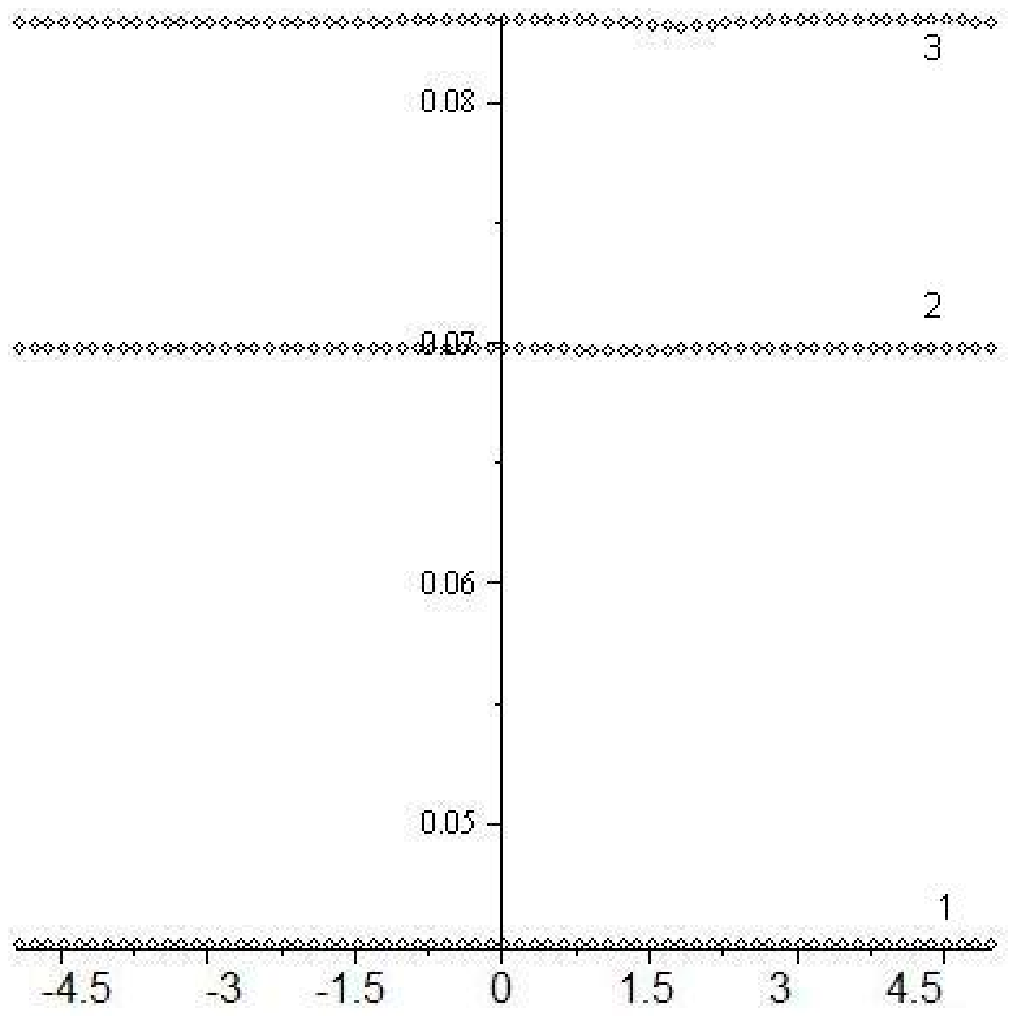}}\label{smilepower}
  \caption{}
\end{minipage}
\end{figure}%

\subsection{Modifications of the Heston model}

Let us analyze the situation when the coefficient $\gamma$  from
equation \eqref{heston2} depends on time. For some interesting cases
of this dependence one can find the Fourier transform of $P(t,f,v)$
and formula for $V(t,f)$. For example, if we set
$\gamma=\frac{1}{T-t}$, then we  get a Brownian bridge-like equation
(see, \cite{Oksendal} describing square of volatility behavior with
start at $v_0 = a > 0$ and end at $v_T = b \geq 0$. Here the
solution will be represented in terms of integrals of Bessel
functions and the solution is cumbersome.

It may seem that the described approach, which helps to find the
conditional  expectation of volatility under fixed returns in the
Heston model, can be successfully applied in other variations of
this model. This is true when initial rate on return has a uniform
distribution. However, this situation is trivial, because the answer
does not contain $f$ and is equal to the expectation of return
obtained from the second equation of model. In the case of
non-uniform initial distribution of return (for instance, Gaussian)
formula \eqref{condexpth} may be non-applicable, even when explicit
expression for $\hat P(t,\mu,\xi)$ can be found. The cause is that
$\hat P(t,\mu,\xi)$ increases as $|\mu|\to\infty$. For example, if
we replace equation \eqref{heston2} with
\begin{equation}\label{heston_mod_2}
 dv_t = -\gamma(v_t - \theta)dt + k
 dW_2,\quad\gamma,\,\theta,\,k>0,
\end{equation}
under initial data \eqref{gf-pzero2}, $\,a=0$, we will get
$$
\hat P(t,\mu,\xi)=\frac{\sqrt{\pi}}{m} \exp\left(\frac
{{k}^{2}t}{{8\gamma^{2}}}{\mu}^{4}\,-\,i\frac {{k}^{2
}t}{{4\gamma^2}}{\mu}^{3}-\left( \frac{\theta \,t}{2}\,+\, \frac
{{k}^{2}t}{8 \gamma^2}\,+\,\frac{1}{4m^2} \right) {\mu}^ {2}\,+\,
i\left( \frac{\theta}{2}-\alpha \right) t\mu\right),
$$
whence it follows that the coefficient of $\mu^4$ in exponent power is positive when $t$ is positive.
This means that integrals from \eqref{condexpth} are divergent.

\section{ Possible application}

Basing on our results one can introduce a rule for estimation of the
company's rating based on stock prices. The natural presumption is
that company's rating increases when return on assets increases and
volatility decreases. Hence for estimation of the company's rating
one can use (very rough) index $R(t,f)=f/V(t,f)$, where $V(t,f)$ is
calculated by formula \eqref{gauss1}. Figs.~3 and ~4 shows the plot
function $R(t,f)$ for three consequent time points for Gaussian and
power distributions, respectively.  Parameters as in Figs.~1 and 2.
We can see that in the Gaussian case the index does not rise
monotonically with return.
\begin{figure}[h]
\begin{minipage}{0.5\columnwidth}
\centerline{\includegraphics[width=0.7\columnwidth]{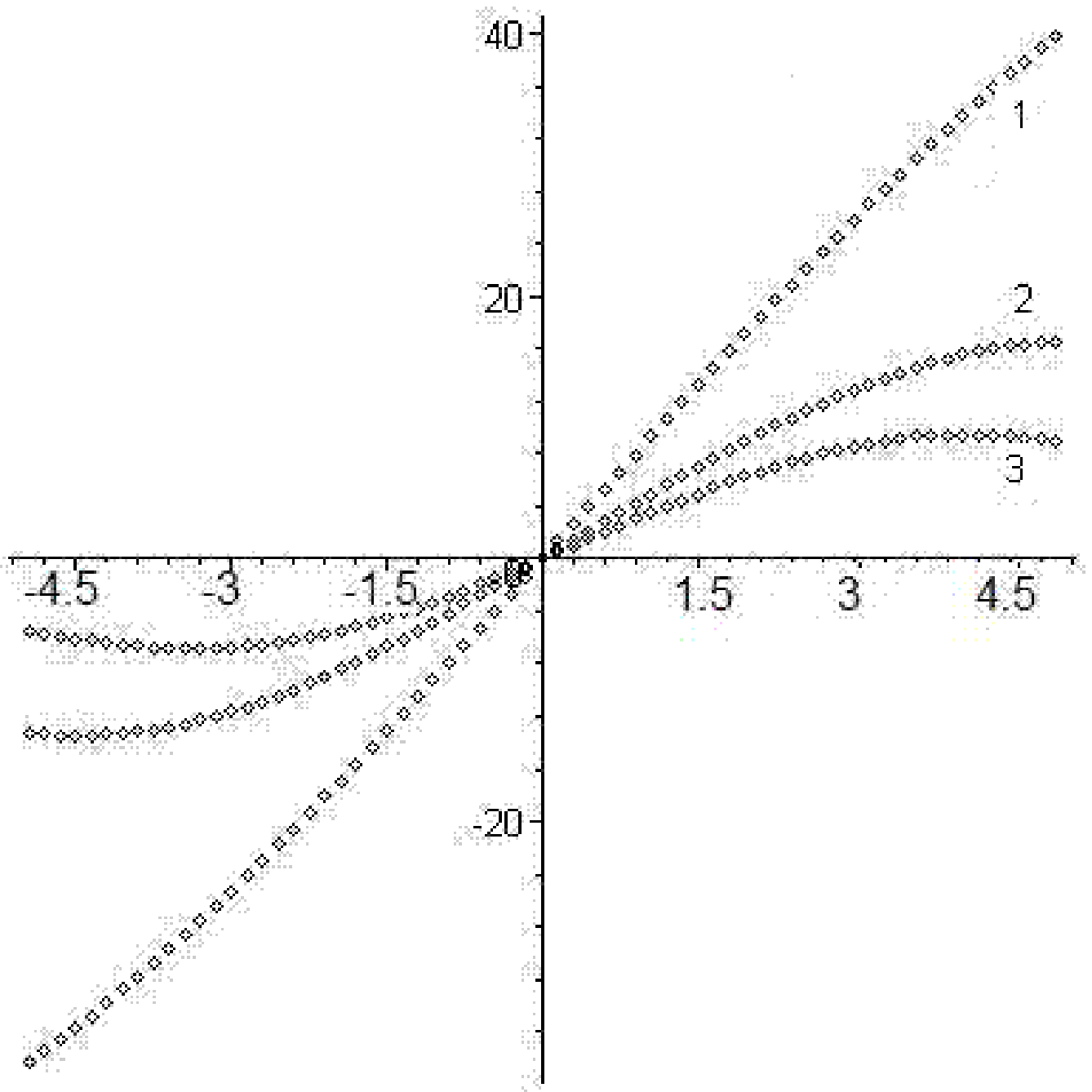}}\label{index}
  \caption{}
\end{minipage}%
\begin{minipage}{0.5\columnwidth}
\centerline{\includegraphics[width=0.7\columnwidth]{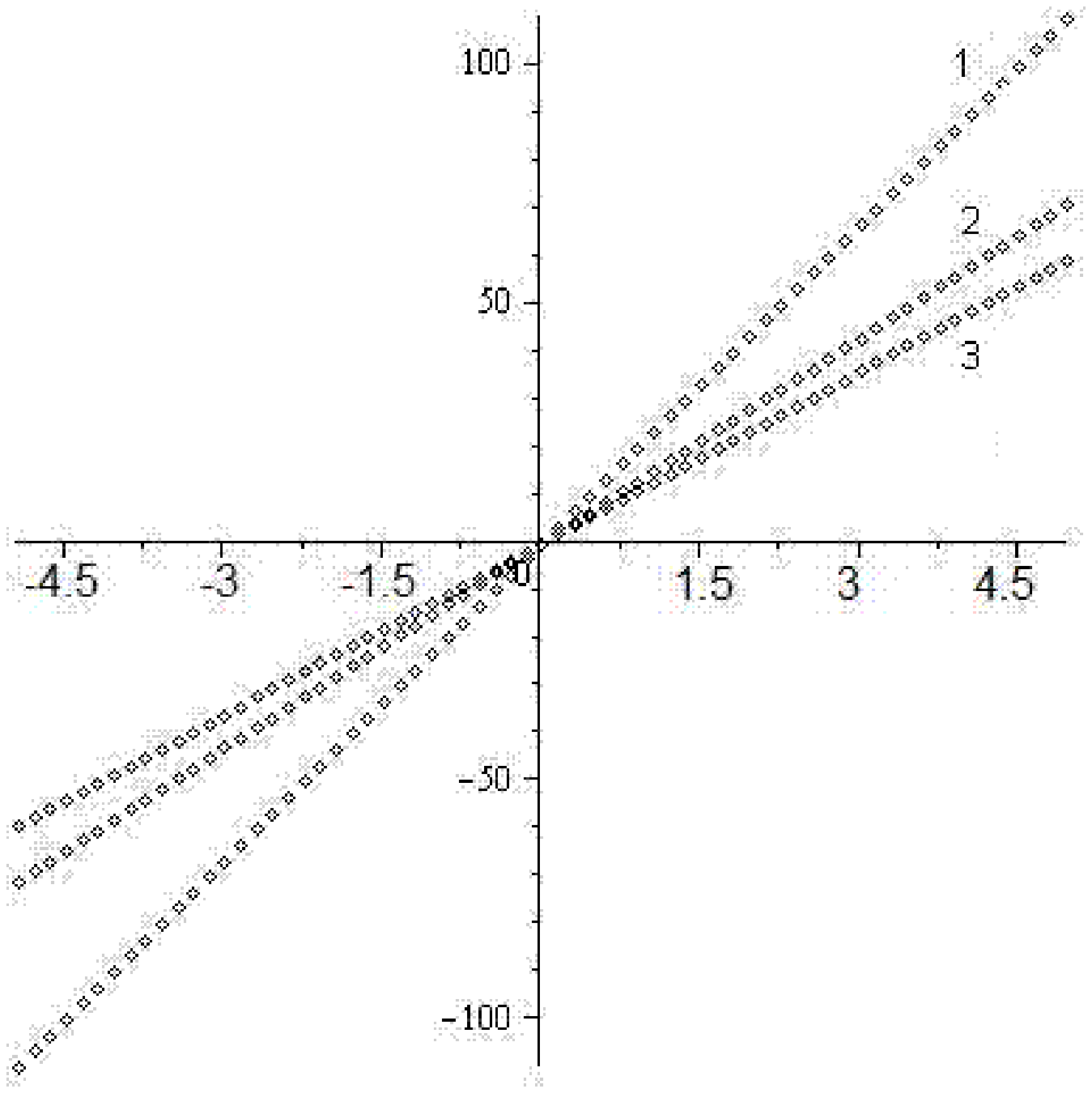}}
\label{indexpower}
  \caption{}
\end{minipage}
\end{figure}

\section{Conclusion and further work} In this article we obtain an estimate
of volatility  given rate on return data in the frame of the Heston
model. This problem has been solved by calculation of the average
volatility under a fixed rate on return and under the supplementary
condition on initial distribution of return and volatility. Namely,
different cases of initial distribution of returns have been
studied: uniform, Gaussian and ``fat-tails'' distributions,
intermediate between them. We revealed that the graph of tthe
averaged volatility is convex downwards  near the mean value of the
stock price return for the Gaussian initial distribution and for
certain distributions decreasing at infinity slower then the
Gaussian one (for which we succeed to find the Fourier transform of
the joint probability density of return and variance explicitly).
For the Gaussian distribution this effect is strong, but it weakens
and becomes negligible as the decay of distribution at infinity
slows down.


Let us note that our formulas can be obtained in a different way,
using the well-known expression for the joint characteristic
function of the log-return and the variance in the Heston model
\cite{Heston} (in the correlated case). This expression was obtained
exploiting the linearity of the coefficients in the respective PDE,
in other word, the fact that the Heston model is affine
\cite{Duffie}. Nevertheless, this way is not  convenient for our
purpose, since it requires an additional integration.


Formulas for the conditional variance at fixed return $V(t,f)$ are
obtained in the present work in the integral form, we compute the
integrals numerically using standard algorithms and  study
asymptotics  of the formulas for small time. The questions on
analysis of  the formulas for larger $t$ and $f$ and on the
asymptotics of $V(t,f)$ as $|f|\to \infty$ and $t\to \infty$ are
open. Moreover, the dependence of  the averaged variance on the
properties of the initial distribution of returns has to be studied
in general case, not only for separate examples, as it was done
here.


\section*{Acknowledgements} This work was supported by the Ministry of
Education of the Russian Federation, project 2.1.1/1399.




\begin{thebibliography}{99}
\bibitem{AR_MMMAS}
S. Albeverio and O.~Rozanova, {\em  The Non-Viscous Burgers Equation
Associated with Random Positions in Coordinate Space: a Threshold
for Blow up Behavior},   Mathematical Models and Methods in Applied
Sciences, 19(2009), pp 1--19.

\bibitem{AR_PAMS}
S. Albeverio and O.~Rozanova, {\em Suppression of Unbounded
Gradients in a SDE Associated with the Burgers Equation}, Proc.
Amer. Math. Soc. 138 (2010), pp. 241--251.


\bibitem{Chorin}
A.\,J. Chorin and O.\,H.~Hald, {\itshape Stochastic Tools in
Mathematics and Science}, Springer, New York, 2006.





\bibitem{Dragulescu} A.A. Dragulescu and  V.M.~Yakovenko,{\em
Probability Distribution of Returns in the Heston Model with
Stochastic Volatility,} { Quantitative Finance}, {2}(2002),
pp.443--453.

\bibitem{Duffie}{D. Duffie, ,D. Filipovi$\rm \acute{c}$ and W.
Schachermayer,}{\em  Affine processes and applications in finance},
 The Annals of Aplied Probability,   { 13}(2003), pp.984--1053.

\bibitem{Feller}{W. Feller, }{\em  Two Singular Diffusion Problems}, {
Annals of Mathematics}, { 54}(1951), pp.173--182.

\bibitem{Fouque}{J.\,P. Fouque,  G.~Papanicolaou, and  K.\,R.~ Sircar},
{\itshape Derivatives in Financial Markets with Stochastic
Volatility}, Cambridge University Press, Cambridge, 2000.


\bibitem{Gatheral} J. Gatheral, {\itshape  The Volatility Surface,} Wiley and Sons, Inc., Hoboken, New Jersey 2006.

\bibitem{Gulisashvili1}{A. Gulisashvili and  E. M. Stein} {\em Asymptotic
Behavior of the Stock Price Distribution Density and Implied
Volatility in Stochastic Volatility Models}, { Mathematical
Finance}, {30} (2010), pp.447--477.

\bibitem{Gulisashvili2}{A. Gulisashvili and  E. M. Stein}, {\em  Asymptotic
Behavior of the Distribution of the Stock Price in Models with
Stochastic Volatility: the Hull-White Model}, { C. R. Acad Sci.
Paris, Ser.I}, { 343}(2006),  pp.519--523.

\bibitem{Heston}
{S.\,L. Heston, } {\em A Closed-Form Solution for Options with
Stochastic Volatility with Applications to Bond and Currency
Options,} { The Review of Financial Studies}, { 6} (1993), pp.
327--343.

\bibitem{Hull}
{J. Hull and  A. White }, {\em  The Pricing of Options on Asset with
Stochastic Volatilities,} { J. Finance}, { 42}(1987), pp.281--300.

\bibitem{Micciche}
{S. Miccich\`e,    G. Bonanno, F. Lillo   and  R. N. Mantegna} {\em
Volatility in Financial Markets: Stochastic Models and Empirical
Results,} { Physica A}, {314}(2002), pp.756-761.

\bibitem{Mitra}
{ S. Mitra, } { em A Review of Volatility and Option Pricing,}
Available at {http://arxiv.org/pdf/0904.1392}

\bibitem{Oksendal}
B. {\O}ksendal,   {\itshape Stochastic Differential Equations: An
Introduction with Applications}, 5th ed.,  Springer, Heidelberg,
2002.

\bibitem{Piessens}
R. Piessens , E. de Doncker-Kapenga, C.\"{U}berhuber  and D.Kahaner,
{\itshape QUADPACK, A Subroutine Package for Automatic Integration}
Springer–Verlag, Berlin, 1983.

\bibitem{Risken}{H. Risken,  } {\itshape The Fokker-Planck Equation. Methods
of solution and applications}, 2ed, Springer, New York, 1989.

\bibitem{Schobel}
{ R.Sch$\rm \ddot{o}$bel, and  J. Zhu, }{\em
  Stochastic
Volatility with an Ornstein - Uhlenbeck Process: An Extension,} {
Europ. Finance Rev.}, { 4}(1999),  pp.23--46.


\bibitem{Scott}
{L.Scott,  }{ \em Option Pricing when the Variance Changes Randomly:
Theory, Estimaton and an Applications,} { J. Finan. Quant. Anal.},
{22}(1987), pp.419--438.

\bibitem{Stein}
{E.M. Stein and  J.C. Stein, } {\em Stock Price Distributions with
Stochastic Volatility: An Analytic Approach}, { Rev. Finan. Stud.},
{4}(1991), pp.727--752.
\end{thebibliography}
\end{document}